\documentclass[manuscript,screen]{acmart}

\AtBeginDocument{%
  }

\setcopyright{acmlicensed}
\copyrightyear{2026}
\acmYear{2026}
\acmDOI{XXXXXXX.XXXXXXX}
\usepackage{wrapfig}

\begin{document}

\title{Carolina Guide: A Multi-Agent RAG System with Institutional Guardrails for Academic Policy Assistance}

\author{Ben Torkian}
\authornote{Corresponding author.}
\email{torkian@email.sc.edu}
\orcid{1234-5678-9012}

\affiliation{%
  \institution{University of South Carolina}
  \city{Columbia}
  \state{South Carolina}
  \country{USA}
}

\author{Jun Zhou}
\affiliation{%
  \institution{University of South Carolina}
  \city{Columbia}
  \country{USA}}
\email{zhouj@mailbox.sc.edu}

% \author{Author 3}
% \affiliation{%
%   \institution{University of South Carolina}
%   \city{Columbia}
%   \country{USA}}
% \email{xxxxx}

% \author{Author 4}
% \affiliation{%
%   \institution{University of South Carolina}
%   \city{Columbia}
%   \country{USA}}
% \email{xxxx}

\renewcommand{\shortauthors}{Torkian et al.}

\begin{abstract}

University students often struggle to navigate complex academic policies, leading to advising bottlenecks and delayed access to critical information. Although large language models (LLMs) offer promise for automated assistance, their tendency toward hallucination and inability to enforce institutional constraints make them unsuitable for high-stakes policy guidance without careful architectural design.
We present Carolina Guide, a retrieval-augmented generation (RAG) system for academic policy assistance at the University of South Carolina (USC). The system employs a modular multi-agent pipeline with institutional guardrails to provide citation-supported, policy-grounded answers to student queries while refusing unsafe requests such as course recommendations or personalized advising. We evaluate the system on a 90-query test set across 6 departments, achieving 98.9\% retrieval success at the rel$\ge$2 threshold (genuinely relevant results) with the first relevant chunk at rank-1 for 98.9\% of queries (MRR@10 for rel $\ge$ 2 = 0.989). Through systematic baseline comparisons and ablation studies, we show that each architectural component—MMR reranking ($\lambda$=0.7), adequate retrieval context (k=20), and citation enforcement—contributes measurable practical value despite limited statistical power at 90 queries. The evaluation of the guardrail on 30 adversarial queries demonstrates Safety F1 of 0.89, correctly refusing 86\% of unsafe queries while maintaining 93\% coverage of benign queries.
These results show that production-ready LLM systems for institutional policy guidance require rethinking standard RAG patterns to prioritize safety, transparency, and departmental autonomy over conversational sophistication.
\end{abstract}



\maketitle

\section{Introduction}

Academic advising is a recurring bottleneck in higher education, driven by complex policies and high advisor workloads. At USC, a 2023 survey found that 72\% of students struggled to find accurate policy information, with email response times of 24–48 hours and office-hour waiting time of 2-7 days. As a result, advisors spend significant time answering routine procedural questions—such as add/drop deadlines or course double-counting—despite the existence of FAQs and handbooks that see low student engagement; students abandoned searches after 3-5 minutes in 43\% of sessions.

LLMs offer a promising interface for natural-language policy assistance, but naive deployments are unsuitable for institutional advising. Our application faces several constraints: 1) \textbf{Hallucination Risk:} LLMs can generate fluent but incorrect guidance, with high-stakes consequences such as registration errors or delayed graduation. 2) \textbf{Safety and Institutional Constraints:} Advising systems must enforce strict boundaries, including refusing personalized academic planning, avoiding speculation about petitions or policy changes, and protecting student data under FERPA.
3) \textbf{Citation and Accountability:} Policy guidance requires explicit provenance so users can verify answers against authoritative documents and institutions can maintain auditability. 4) \textbf{Departmental Autonomy and Maintenance:} Policies are owned by individual departments and change frequently; systems must allow non-technical staff to update content with zero downtime while enforcing role-based access control.
These constraints define a system design problem rather than a purely NLP task. While existing retrieval-augmented generation (RAG)-based implementations can reduce hallucinations. However, they prioritize conversational fluency and retrieval recall over institutional deployment, which requires the deterministic safety guaranties, citation transparency, and operational sustainability required for institutional deployment.
   
Considering these constraints, we present Carolina Guide, an RAG-based multi-agent assistant that answers policy questions with cited sources while enforcing institutional guardrails to refuse unsafe requests (course recommendations, speculation, PII).
% we present Carolina Guide, a production-ready system for university policy assistance that prioritizes institutional compliance and operational sustainability over conversational sophistication. 
% Our core contribution is a holistic architecture integrating five components optimized for high-stakes advising:
Our system has five components optimized for high-stakes advising:
\textbf{1) Five-Agent Sequential Pipeline:} We propose a sequential multi-agent architecture that decomposes query processing into specialized roles, to enable deterministic safety enforcement and auditability. Unlike single-prompt RAG systems that rely on prompt engineering, our design assigns intent handling, retrieval, answer generation, policy enforcement, and response composition to distinct agents coordinated by an orchestrator (Section~\ref{sec:method}).    
\textbf{2) Relational-First Hybrid Database:} 
% We adopt a relational-first hybrid storage architecture that treats the relational database as the system of record and the vector database as a derived semantic index. Unlike vector-first RAG designs, this approach enables transactional guarantees for versioning, role-based access control, and auditability, while still supporting efficient semantic retrieval for natural language queries (Section~\ref{sec:method}).
We use a relational-first hybrid architecture in which the relational database serves as the system of record and the vector store as a derived semantic index. This design supports versioning, role-based access control, and auditability while preserving efficient semantic retrieval (Section~\ref{sec:method}).
\textbf{3) Document Indexing Pipeline:}   
Department administrators publish policy documents through a web portal without engineering involvement. The system automatically detects changes, reindexes content with section-aware chunking, generates embeddings, and updates the Qdrant vector store with monitoring and error recovery. 
% This pipeline supports domain management by domain experts, not technical specialists. 
\textbf{4) Hard Guardrails with Explicit Refusals:} 
 Rather than relying on training or soft safety constraints, we enforce guardrails architecturally. The Orchestrator agent blocks prohibited queries (e.g., course recommendations, speculation attempts, PII queries) before retrieval, while the Policy Guardrail agent validates generated responses and rewrites or rejects violations post hoc. This defense-in-depth design ensures consistent and institutionally trustworthy behavior.
 \textbf{5) Citation as First-Class Requirement:}    
 % All substantive answers require explicit citations to source documents with title, section, etc. The Response Composer extracts and formats citation metadata from retrieved chunks. The system then enforces a citation threshold—answers without proper grounding trigger fallback responses or escalation to human advisors.
 % This design prioritizes verifiability and institutional accountability over conversational fluency.
 All substantive answers must include explicit citations to the source documents (e.g., title and section). The response composer pulls citation metadata directly from the retrieved content and formats it for display. If an answer cannot be properly grounded, the system falls back to a refusal or escalates the query to a human advisor. This design favors verifiability and institutional accountability over conversational polish.

 % This system balances the need for a natural student interface with the strict safety and autonomy requirements of a large institution. Our evaluation proves that this balance is possible: the system provides accurate, cited policy help while keeping control in the hands of the departments. By design, we avoided integrating sensitive student records or personalized recommendations, focusing instead on safe, verifiable operation. Evaluation of 90 queries in six departments at USC showed strong retrieval quality, with a success rate $98.9\%$ at the $rel \ge 2$ threshold and a Safety F1 of $0.89$. While this study focuses on the immediate metrics of accuracy and latency, it lays the groundwork for future research into long-term effects on advisor workloads and student outcomes.

 The system is designed to balance a natural, student-facing interface with the strict safety and autonomy requirements of a large institution. Our evaluation shows that this balance is achievable: the system delivers accurate, well-cited policy guidance while keeping content control within individual departments. By design, we exclude sensitive student records and personalized recommendations, focusing instead on safe and verifiable operation. Across 90 queries in six departments at USC, the system achieves strong retrieval performance, with a 98.9\% success rate at the rel $\ge$ 2 threshold and a Safety F1 of 0.89. While this study focuses on accuracy and latency, it also establishes a foundation for future work on long-term impacts to advisor workload and student outcomes.

\section{Related Work}
\label{sec:related_work}

Recent advances in artificial intelligence have driven progress across many domains~\cite{liu2020ensemble, zhou2019framework}.
%lung2025ensemble, 
In education, LLMs have renewed interest in intelligent advising systems. However, institutional policy advising introduces requirements for safety, accuracy, provenance, and organizational control that exceed those addressed by most conversational AI systems. Our work integrates insights from RAG, multi-agent LLM architectures, guardrails, hybrid data management, and educational AI.

\textbf{Retrieval-augmented generation} mitigates hallucination by grounding generation in external knowledge~\cite{torkian2026ifs}. Foundational work combines dense retrieval with generation~\cite{lewis2020retrieval}, with subsequent advances in sparse, dense, and hybrid retrieval~\cite{robertson2009probabilistic, karpukhin2020dense, lin2021batch}, as well as iterative retrieval and query decomposition~\cite{shao2023enhancing, khattab2022demonstrate}. While RAG has been widely applied to tutoring and learning materials, institutional advising requires stricter guarantees: explicit citation, metadata-aware filtering, and zero tolerance for factual error. Existing systems often prioritize recall and fluency, whereas policy guidance demands verifiability.

\textbf{Multi-agent LLM systems} decompose complex tasks through collaboration or coordination~\cite{wu2024autogen}
%park2023generative, wu2024autogen}
. These approaches emphasize flexibility and emergent behavior. In contrast, our system adopts a deterministic, sequential pipeline in which agents perform fixed operational roles (intent handling, retrieval, generation, policy enforcement, response composition). This design prioritizes predictability, auditability, and institutional control over adaptive coordination, aligning with emerging critiques of unconstrained agent collaboration in high-stakes settings~\cite{torkian2026transcript}.

\textbf{Guardrails and safety} mechanisms have been explored through behavioral testing~\cite{ribeiro2020beyond}, RLHF~\cite{ouyang2022training}, and constitutional AI~\cite{bai2022constitutional}. These methods primarily target harmful or toxic content. Institutional advising introduces distinct constraints, such as preventing personalized recommendations, speculative policy interpretation, and unauthorized data use. Our work aligns with approaches that enforce safety architecturally, using explicit guardrail components rather than prompt- or training-based controls.

\textbf{Managing institutional knowledge} requires both transactional guarantees and semantic retrieval. Relational databases support access control, versioning, and auditability~\cite{date1994introduction}, while vector databases enable large-scale similarity search~\cite{johnson2019billion}. Most RAG systems adopt vector-first designs. Our relational-first hybrid approach, grounded in established data engineering principles~\cite{kleppmann2019designing}, treats the relational database as the system of record and the vector store as a derived index. This enables department-scoped ownership, version-controlled updates, and metadata-constrained retrieval—capabilities critical for multi-department deployments.

Finally, prior work on \textbf{educational AI assistants} focuses primarily on pedagogical support~\cite{kerlyl2006bringing, goel2018jill}. Policy advising differs fundamentally: it requires authoritative sourcing, avoidance of personalization, and strict adherence to institutional governance. Similarly, learned citation generation~\cite{menick2022teaching, gao2023enabling} can hallucinate sources, whereas our system enforces provenance through deterministic metadata propagation.

Overall, while prior work addresses individual components—retrieval, agents, guardrails, or knowledge management—few systems integrate them into a production-ready, department-owned architecture for institutional policy advising. Our work demonstrates that such settings require rethinking standard RAG designs to prioritize safety, provenance, and operational sustainability over conversational sophistication.

\section{System Design and Implementation}
\label{sec:method}
\subsection{System Architecture}
\begin{figure}[htp]
  \centering
  \includegraphics[width=0.9\linewidth]{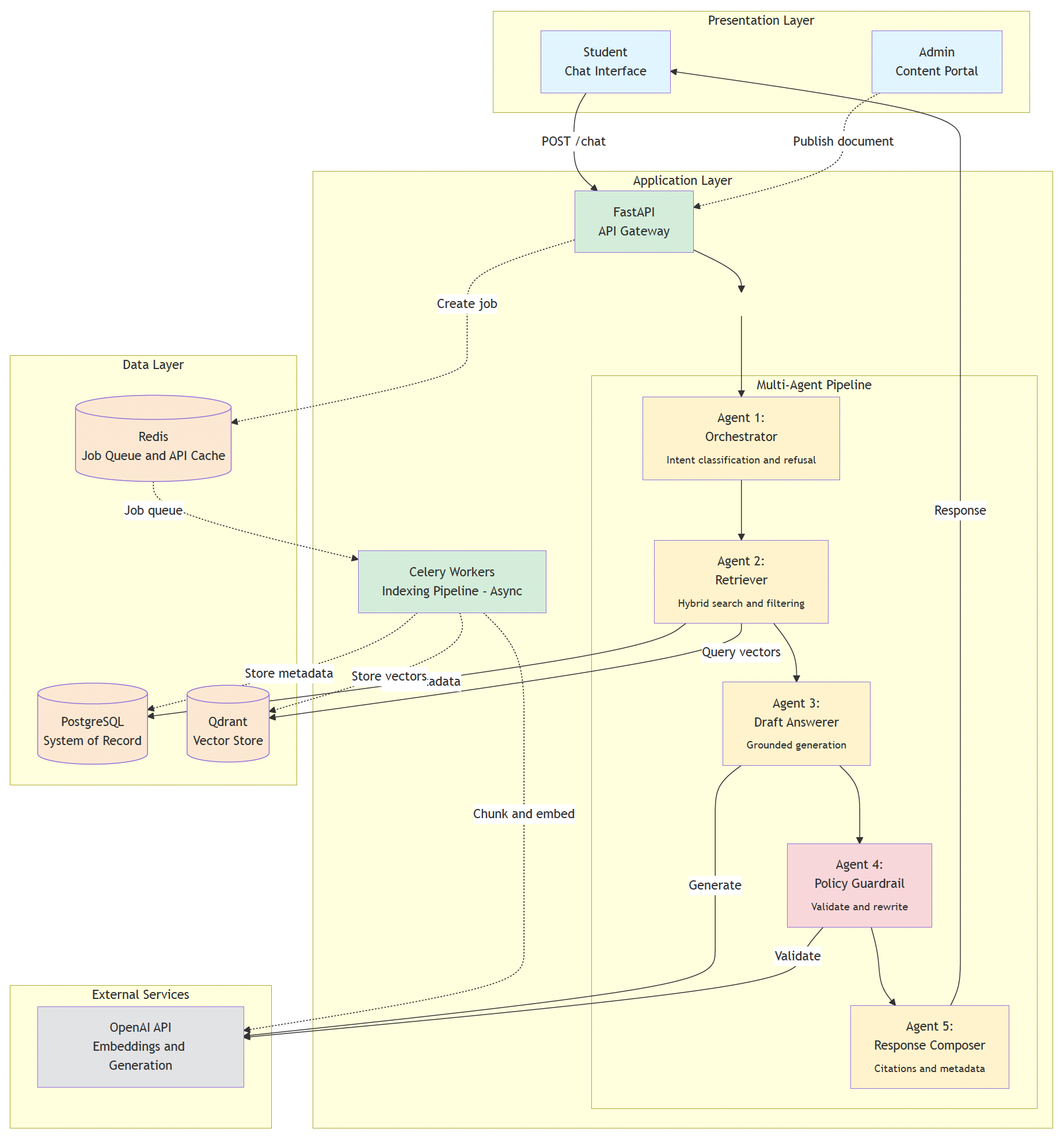}
  \caption{System architecture showing three-layer design with hybrid PostgreSQL-Qdrant database, five-agent pipeline, and asynchronous indexing. Solid arrows: synchronous (query path), dashed arrows: asynchronous (indexing).}
  \label{fig:ADV_architecture}
  \vspace{-1em}
\end{figure}

The Carolina Guide implements a multi-agent RAG architecture designed to provide policy-compliant advising information to university students. The system (Fig.~\ref{fig:ADV_architecture}) comprises three primary layers—presentation, application, and data.   

The presentation layer consists of two web-based interfaces built with Next.js 14: a student-facing chat interface for query submission and a department administrator portal for knowledge base management. Both interfaces communicate with the   application layer via RESTful APIs secured with JWT-based authentication.   

% Agent 1: Orchestrator classifies query intent and detects department scope.
% 1
% Agent 2: Retriever performs vector search (Qdrant HNSW) with MMR
% reranking (￿=0.7).
% Agent 3: Draft Answerer generates grounded responses with citation markers.
% Prompt: “Answer using ONLY the provided context.”
% Agent 4: Policy Guardrail targets violations (course recommendations, speculation,
% PII, cross-campus mismatches; known limitation: cross-campus detection
% currently limited). If violation detected, overrides draft with refusal
% template.
% Agent 5: Response Composer formats structured citations.
% Orchestration: Sequential pipeline—Query → Orchestrator →

The application layer implements a FastAPI backend 
%(Python 3.11) 
% that orchestrates multi-agent query processing, document   management, and access control. This layer contains a five-Agent sequential pipeline: the Orchestrator performs intent   classification and early refusal of prohibited queries. The Retriever executes hybrid semantic search with metadata filtering  over authorized content. The Draft Answerer generates initial responses grounded in the retrieved context with structured   citation requirements. The Policy Guardrail agent enforces institutional constraints through automated content analysis and   rewriting. The Response Composer also assembles final output with citation metadata, confidence scores, and escalation flags. 
that orchestrates multi-agent query processing, document management, and access control. This layer contains our core five-Agent sequential pipeline: the Orchestrator performs query intent classification and detects department scope. The Retriever executes hybrid semantic search with metadata filtering  over authorized content. The Draft Answerer generates grounded responses with in the retrieved citation markers.  The Policy Guardrail agent targets violations (course recommendations, speculation, PII, cross-campus mismatches; known limitation: cross-campus detection currently limited). If violation detected, overrides draft with refusal template. The Response Composer formats structured citations. 
This application layer maintains strict separation between synchronous request handling (API endpoints) and asynchronous background processing (indexing pipeline). The synchronous path handles authentication, input validation, agent orchestration, and response assembly
%with a target latency of <2 seconds for 95th percentile queries
. The asynchronous path processes document indexing tasks through a distributed task queue architecture using Celery 5.3 with Redis 7.2 as the message broker.  

% The data layer employs a polyglot persistence strategy optimized for distinct access patterns. PostgreSQL 15 serves as the system of record, storing policy documents, content versions, department ownership, role-based access control (RBAC), indexing job state, and audit logs. All authoritative policy content and metadata reside exclusively in PostgreSQL, ensuring transactional consistency and traceability.
% Qdrant 1.7 functions as a derived semantic index, storing vector embeddings generated from published documents to support semantic retrieval. Qdrant is asynchronously synchronized from PostgreSQL through a Celery-based embedding pipeline, ensuring that vector data are never authoritative and can be regenerated deterministically from the relational source. Metadata filtering (e.g., department scope, publication status, document version) is enforced at query time to maintain isolation between departmental knowledge bases. Redis provides both job queue persistence and session caching for the API layer.  
The data layer uses a polyglot persistence strategy tailored to specific access patterns. PostgreSQL 15 serves as the system of record, managing policy documents, content versioning, department, role-based access control (RBAC), indexing job status, and audit logs to ensure transactional consistency.
Qdrant 1.7 serves as a derived semantic index, populated asynchronously via a Celery-based pipeline. Because Qdrant is treated as a secondary index, vector data can be regenerated deterministically from PostgreSQL at any time. To maintain isolation between departments, metadata filters (e.g., department scope and document status) are enforced at the query level. Redis handles session caching and job queue persistence for the API layer.

All vector embeddings are generated using OpenAI's text-embedding-3-small model (1536 dimensions)
, selected for its   balance of semantic fidelity and computational efficiency. Answer generation employs GPT-4 Turbo (gpt-4-turbo-2024-04-09)   with structured output formatting via function calls to ensure consistent citation formatting and guardrail enforcement.
\subsection{Data Model and Schema Design}
The relational schema implements a department-scoped knowledge base with granular role-based access control and full auditability. Academic and administrative units are represented as departments, each maintaining an isolated policy repository to prevent cross-department information leakage. User accounts are managed with two privilege levels—department administrators and super administrators—while a junction table enables controlled access to multiple departments when required (e.g., interdisciplinary programs).

Policy documents are stored with metadata including department association, effective term, source URL, and status, and follow a two-stage lifecycle in which draft documents are editable but excluded from retrieval. Upon publication, documents are versioned and trigger asynchronous indexing jobs. Document chunks are tracked relationally with positional and token metadata, while full embeddings are stored in Qdrant. Indexing jobs are monitored for status and errors, and all data modifications are recorded in an audit log for compliance. Vector metadata mirrors relational constraints—such as department, version, and publication status—to ensure filtered retrieval and prevent draft content from appearing in query results.

\subsection{Document Indexing Pipeline.} 
Published policy documents are transformed into semantically searchable vectors through an asynchronous, multi-stage indexing pipeline triggered when a department administrator publishes a document via the admin portal. Upon publication, the API atomically updates the document status, increments its version, creates a job record with status queued, and enqueues a task to Redis, allowing the API request to complete quickly (<200 ms) while deferring embedding generation to background workers.

Celery workers claim queued jobs and execute a deterministic pipeline consisting of: (1) input validation and encoding checks, (2) section-aware chunking using a sliding window (target $\sim$600 tokens with 100-token overlap) with section headers prepended, (3) batched embedding generation using OpenAI’s text-embedding-3-small model with rate-limit backoff and jitter, (4) deterministic vector ID construction (document id · version · chunk index), and (5) atomic vector upsert into Qdrant using the batch API with wait set to true.

To ensure production robustness, the pipeline enforces a two-phase commit pattern in which relational chunk records are created within a transaction that is committed only after a successful Qdrant upsert; failures trigger rollback and automated cleanup of orphaned vectors. Additional safeguards include embedding validation (dimension and NaN/Inf checks), idempotent unique constraints on (document id, chunk index) to support safe retries, exponential-backoff handling of transient API errors, and a nightly garbage-collection task for residual vectors. This design preserves PostgreSQL as the system of record while maintaining a consistent, low-latency semantic index.

\subsection{Multi-Agent Query Orchestration}
Query processing uses a sequential multi-agent architecture in which specialized agents perform distinct reasoning steps with explicit handoffs.
This design favors interpretability and policy compliance over end-to-end optimization, enabling fine-grained behavioral control and complete auditability.
Upon query submission, a central orchestrator coordinates a synchronous five-agent pipeline, with each agent’s output serving as input to the next. The Orchestrator classifies intent, determines departmental scope, and refuses prohibited queries. The Retriever performs semantic search over published content with metadata filtering and Maximal Marginal Relevance (MMR)~\cite{carbonell1998use} reranking ($\lambda=0.7$). The Draft Answerer generates a grounded response from retrieved context with structured sections and inline citations. The Policy Guardrail validates the draft against institutional constraints, applying automated fixes or rejecting noncompliant outputs. Finally, the Response Composer assembles the final response with full citation metadata, confidence scores, and escalation flags. All intermediate decisions and outputs are logged to provide end-to-end traceability and support auditing and debugging.

\subsection{Guardrail Implementation and Safety Mechanisms}
The system employs a defense-in-depth safety architecture with enforcement at three stages. Pre-flight input validation is performed by the Orchestrator, which detects prohibited query types and issues immediate, templated refusals before retrieval occurs. During generation, prompt-level constraints and structured outputs via GPT-4 Turbo function calling reduce unsafe behaviors. After generation, a Policy Guardrail analyzes the draft response: minor violations are automatically corrected, while major violations trigger rejection. Additional safeguards include a minimum citation requirement for substantive answers, FERPA-compliant handling of user input (no PII sent to external LLMs), and comprehensive safety logging of all guardrail events for auditability and continuous improvement.

% % \textbf{Smart Query Suggestions for Ambiguous Rejections}

Strict guardrails often create dead-end refusals when queries contain out-of-scope trigger terms but reflect valid policy intent. To mitigate this, the system introduces smart query suggestions when a query is rejected as out-of-scope. The Orchestrator extracts domain-relevant keywords, checks for related in-scope content using PostgreSQL full-text search, and—if matches exist—generates alternative, policy-compliant query phrasings. These suggestions are presented in the UI as clickable options that resubmit the rephrased query automatically.

This approach transforms rigid refusals into guided exploration, improving user experience without relaxing safety boundaries. While the mechanism incurs modest additional latency ($\sim$150 ms when activated), it avoids speculative answers and preserves institutional constraints. The design favors helpfulness over strict rejection in ambiguous cases, with empirical validation left to future user studies examining click-through rates, satisfaction, and false-positive suggestions.

\section{Evaluation Design and Protocol}
\label{sec:evaluation}

% \textbf{Evaluation Protocol.} 
We evaluate the system using a controlled experimental protocol designed to assess retrieval quality and robustness under production-like conditions. All experiments use GPT-4o-mini with temperature set to 0.0 to eliminate stochastic variation. The evaluation is conducted on a test set of 90 queries across six academic departments with 16 policy documents (264 chunks). Retrieval uses top-k = 20 candidates with a similarity threshold of 0.15, followed by MMR reranking ($\lambda$ = 0.7) to select the top 10 results. The system enforces a minimum of one citation per response.
%, processes queries sequentially, and was evaluated on February 3, 2026.

\textbf{Similarity Threshold Selection.}
The similarity threshold (0.15) was chosen via empirical calibration on a development set to balance recall and noise. Queries with no candidates above this threshold result in a system refusal and referral to human advising.

% The similarity threshold (0.15) was selected through empirical calibration on a development set. Lower thresholds increase recall by admitting more candidate chunks, while higher thresholds reduce noise at the cost of coverage. A threshold of 0.15 was found to balance these tradeoffs effectively. Queries for which no chunks exceed this threshold result in a system refusal with referral to human advising.

\textbf{Ground Truth Labeling.}
% Retrieval relevance was assessed through expert annotation. A domain expert (USC advisor with over five years of experience) labeled the top-10 retrieved chunks per query, yielding 900 total annotations. Chunks were rated on a four-point ordinal scale:
% 0 — not relevant;
% 1 — marginally relevant;
% 2 — relevant;
% 3 — highly relevant.
Retrieval relevance was annotated by a domain expert (USC advisor, >5 years experience). The top-10 chunks per query were labeled (900 total annotations) on a four-point ordinal scale: 0 (not relevant), 1 (marginal), 2 (relevant), and 3 (highly relevant).

\textbf{Success Criteria and Metrics.} 
We report success at multiple relevance thresholds to capture different notions of correctness: Success(rel $\ge$ 1) includes marginally relevant results; Success(rel $\ge$ 2) considers only genuinely relevant results and serves as the primary metric; Success(rel = 3) reflects the strictest criterion.
Retrieval performance is measured using:
(1) Relevant Chunks@k, the average number of chunks with relevance above a threshold in the top-k;
(2) MRR@k, the mean reciprocal rank of the first chunk meeting the relevance threshold; and
(3) nDCG@k, which captures graded ranking quality using exponential gain g(rel) = $2^{rel}-1$.
MRR and nDCG~\cite{jarvelin2002cumulated} differ in sensitivity: MRR reflects only the position of the first relevant chunk under a binary threshold, whereas nDCG accounts for graded relevance across all retrieved positions. In our data, the rank-1 relevance distribution is dominated by rel = 2 and rel = 3 chunks (mean relevance 2.10), explaining the combination of MRR@10 = 1.0 with a lower nDCG@10 = 0.548.

\textbf{Implementation Details and Variance.}
Documents are chunked into segments ranging from 100 to 800 tokens (mean $\approx$ 400) with 80–120 token overlap, reflecting semantic structure: shorter chunks preserve headers and brief statements, while longer chunks retain complete policy explanations. Vector search is implemented using Qdrant HNSW (m = 16, ef\_construction = 100, ef\_search = 100), with embeddings generated by text-embedding-3-small (1536 dimensions). Ideal DCG is computed using only judged top-10 chunks, with unjudged items treated as non-relevant.
Across repeated runs, query-level retrieval variance remained below 1\%, likely attributable to approximate nearest-neighbor search behavior, index state changes, or external embedding service variability.

\section{Results}
\subsection{System Performance}
Table~\ref{tab:ADV_latency_capacity}. shows the system's performance profile and capacity benchmarks. We observed a mean end-to-end latency of 2.8s with a percentile of 95\% of 3.2s, comfortably within our <5s design target for interactive sessions. Retrieval overhead is negligible, consistently clocking under 200ms; this efficiency is a direct result of Qdrant’s HNSW indexing paired with pre-retrieval metadata filtering.
As expected, the primary bottleneck is LLM inference, which accounts for the vast majority of the execution trace. Current single-instance throughput is approximately 21 RPM (queries per minute). While the system is currently compute-bound by the LLM, the stateless nature of our backend service allows for horizontal scaling by deploying additional instances against the shared PostgreSQL and Qdrant cluster.
\begin{table}[htp]
  \caption{Latency and Capacity}
  \label{tab:ADV_latency_capacity}
  \begin{tabular}{lll}
    \toprule
    Metric &Value& Notes\\
    \midrule
    Mean Response Time & 2.8s & p95: 3.2s\\
Retrieval Latency & <200ms & Qdrant HNSW\\
Sequential Throughput & $\sim$21 queries/min & LLM-constrainedt\\
Total Chunks Indexed & 264 & 16 documents\\
  \bottomrule
  \vspace{-1em}
\end{tabular}
\end{table}

\subsection{RAG Quality}
% Table \ref{tab:ADV_retrievalaccuracy} reports retrieval performance over 90 queries under increasingly strict relevance criteria. Under the primary setting (rel $\ge$ 2), the system successfully retrieves a genuinely relevant chunk within the top-10 for 98.9\% of queries (89/90), with an MRR@10 of 0.989, indicating that relevant content is almost always ranked first. Precision at rank 1 is likewise 98.9\%, and the average number of relevant chunks in the top-10 is 1.24, reflecting focused retrieval rather than over-retrieval. Performance reaches 100\% success and MRR under the lenient threshold (rel $\ge$ 1). In contrast, enforcing the strictest criterion (rel = 3) yields substantially lower scores (13.3\% success), highlighting that highly specific, fully satisfactory answers are rare and underscoring the sensitivity of retrieval metrics to relevance definitions.
Table~\ref{tab:ADV_retrievalaccuracy}. shows our retrieval performance across three relevance definitions. Under the primary criterion (rel $\ge$ 2), the system consistently retrieves genuinely relevant policy content near the top of the ranking, with minimal over-retrieval. Performance is near ceiling under the more lenient threshold, while the sharp drop under the strictest criterion (rel = 3) reflects the scarcity of fully exhaustive answers rather than a failure to retrieve useful content. Together, these results indicate that the RAG pipeline is well calibrated for policy advising, where reliably surfacing at least one clearly relevant source is more important than optimizing for rare, fully comprehensive matches.
\begin{table}[htp]
  \caption{Retrieval Quality at Multiple Relevance Thresholds}
  \label{tab:ADV_retrievalaccuracy}
  \begin{tabular}{lllll}
    \toprule
    Threshold &Success @10 &MRR@10 &Rel. Chunks@10 &Top-1 Precision\\
    \midrule
rel$\ge$2 (stricter)& 98.9\% & 0.989 & 1.244 & 98.9\%\\
rel$\ge$1 (lenient) &100.0\% & 1.000 & 1.254 & 100.0\%\\
rel=3 (strictest) &13.3\% & 0.133 & 0.133 & 13.3\%\\
  \bottomrule
   \end{tabular}
 \end{table}
\subsection{Baseline Comparisons}
%Baseline comparison and statistical analysis. Table \ref{tab:baseline-comparison} compares the full system against ablated variants under the primary relevance criterion (rel ≥ 2). The full configuration achieves the highest success rate, with 89/90 queries successfully retrieving a genuinely relevant result. A McNemar’s exact two-sided test comparing the full system to the No-MMR variant (87/90 successes) yields p = 0.50, indicating that the observed 2.2 percentage point improvement is not statistically significant at α = 0.05 (95% CI: −3.5% to +7.7%, Wilson score). This lack of significance is expected given the high baseline performance (>96%) and limited sample size (n = 90), which constrain statistical power. Nonetheless, all observed differences favor the full system, and the practical impact remains meaningful: at a deployment scale of 10,000 queries annually, the improvement corresponds to approximately 220 fewer retrieval failures. Component-wise, MMR reranking contributes +2.2 pp in success rate, adequate context contributes +1.1 pp, and citation enforcement contributes +1.7 pp (see Section 4.4). While disabling MMR (λ = 1.0) slightly improves nDCG@10 (0.552 vs. 0.548), this reflects a trade-off between ranking quality and retrieval success; we prioritize success rate to minimize end-to-end retrieval failures.

As shown in Table \ref{tab:baseline-comparison}., the full system achieves the highest success rate under the rel $\ge$ 2 criterion. The difference relative to the No-MMR variant (2.2 pp) is not statistically significant by McNemar’s exact test (p = 0.50; $\alpha$ = 0.05), reflecting limited power due to near-ceiling performance on a 90-query test set. However, the consistent degradation across all ablations suggests that MMR reranking, and adequate context
% , and citation enforcement 
each contributes positively to robust retrieval, even when gains in nDCG are marginal or mixed.
\begin{table}[htp]
\centering
% \caption{Baseline Comparison on 90-Query Test Set (Citation enforcement baseline uses 60-query stratified subset as Table~\ref{tab:citation-ablation}.)}
\caption{Baseline Comparison}
\label{tab:baseline-comparison}
\begin{tabular}{lcccc}
\hline
Approach & Success (rel $\ge$ 2) & MRR@10 & Rel. Chunks@10 & nDCG@10 \\
\hline
Carolina Guide (Full) & 98.9\% & 0.989 & 1.244 & 0.548 \\
No MMR Reranking ($\lambda$ = 1.0) & 96.7\% & 0.967 & 1.267 & 0.552 \\
Minimal Context (k=5) & 97.8\% & 0.978 & 1.256 & 0.558 \\
% No Citation Enforcement & 98.3\% & 1.250 & 0.983 & 0.529 \\
\hline
\end{tabular}
\end{table}
\subsection{Ablation Studies}
\subsubsection{MMR Lambda}
Table \ref{tab:mmr-ablation} shows that MMR with $\lambda = 0.7$ achieves the peak success rate ($98.9\%$), beating both pure diversity and pure relevance. While $\lambda = 1.0$ slightly improves nDCG@10, we opted for $0.7$ to prioritize overall success and minimize retrieval failures.
\begin{table}[htp]
\centering
\caption{MMR Ablation (rel $\ge$ 2 threshold)}
\label{tab:mmr-ablation}
\begin{tabular}{llll}
\hline
$\lambda$ & Success & nDCG@10 & Interpretation \\
\hline
0.0 & 97.8\% & 0.545 & Pure diversity \\
0.7 (production) & 98.9\% & 0.548 & Maximizes success rate\\
1.0 & 96.7\% & 0.552 & Pure relevance \\
\hline
\end{tabular}
\end{table}

\subsubsection{Retrieval Context Size Ablation.}
As shown in Table~\ref{tab:retrieval-k-ablation}, retrieval success improves from $k=5$ to $k=20$ and then plateaus. Increasing to $k=100$ provides no additional success gain while incurring a fivefold cost, making $k=20$ an efficient quality–cost trade-off.
\begin{table}[htp]
\centering
\caption{Retrieval Context Size Ablation}
\label{tab:retrieval-k-ablation}
\begin{tabular}{lll}
\hline
$k$ & Success (rel $\ge$ 2) & Cost\\
\hline
5 & 97.8\% & 0.25x \\
20 (production) & 98.9\% & 1x \\
100 & 98.9\% & 5x \\
\hline
\end{tabular}
\end{table}

\subsubsection{Citation Enforcement Ablation.}
\begin{table}[h]
\centering
\caption{Citation Threshold Ablation (60-query stratified set)}
\label{tab:citation-ablation}
\begin{tabular}{lccccc}
\hline
\textbf{Min Citations} & Success & Refusal Rate & Unsupported Claims (Manual Review) \\
\hline
0 (none) & 96.7\% (58/60) &  5.0\% & 10\% (2/20 reviewed)\\
1 (production) & 100\% (60/60) & 6.7\% & 0\% (0/20 reviewed)\\
3+ & 100\%  (60/60) & 10.0\% & 0\% (0/20 reviewed) \\
\hline
\end{tabular}
\end{table}
A manual review of 20 responses per condition by two annotators shows that requiring at least one citation eliminates unsupported claims (10\% → 0\%) while achieving perfect success on the 60-query subset. A Fisher’s exact test on the review labels is not statistically significant (p = 0.49; n = 40), reflecting limited power, but the descriptive results support a minimum citation threshold as an effective grounding constraint.

\subsection{Guardrail Effectiveness}
As shown in Table~\ref{tab:guardrail-effectiveness}, the guardrail~\cite{bai2022constitutional} achieves strong overall performance on 30 adversarial queries (precision 92\%, recall 86\%, F1 = 0.89). The single cross-campus failure reflects a known limitation where a USC Aiken query retrieved USC Columbia content and was not flagged. One out-of-scope query is excluded from metrics due to a natural refusal caused by retrieval failure rather than a guardrail decision.
\begin{table}[htp]
\centering
\caption{Guardrail Performance (30 Adversarial Queries)}
\label{tab:guardrail-effectiveness}
\begin{tabular}{llll}
\hline
Query Type & Count & Correct Refusals &Coverage\\
\hline
Unsafe Course Rec & 5 & 5/5 (100\%) & -\\
Unsafe Speculation & 5 & 4/5 (80\%) & -\\
Unsafe PII & 3 & 3/3 (100\%) & -\\
Cross-Campus Error & 1 & 0/1 (0\%)† & -\\
Out-of-Scope (no content) & 1 & N/A & N/A‡\\
Benign & 15  & - & 14/15 (93\%)\\
Total & 30 & 12/14 (86\%) & F1=0.89\\
\hline
\end{tabular}
\vspace{-1em}
\end{table}
\subsection{Integration Testing}
We evaluated the system using 12 end-to-end integration tests spanning conversational behavior (3/3), department routing (2/2), public endpoint access (3/3), and citation quality (2/2). All tests passed, and no hallucinated sources were observed.

\subsection{Limitations}
The evaluation is constrained by limited statistical power: with 90 queries and high baseline performance (>96\%), 1–2 pp differences do not reach significance (McNemar’s p > 0.05); approximately 600 queries would be required for 80\% power. The test set was developed iteratively alongside tuning, though a partial post-tuning validation of 40 additional queries achieved comparable performance; fully independent test sets are recommended. Measurement of unsupported claims is based on a small manual review (40 responses), yielding insufficient power (Fisher’s p = 0.49). Guardrail evaluation is limited to 30 single-turn adversarial queries and does not cover prompt injection or multi-turn attacks. Finally, results are drawn from a single institution with a limited document set, and broader generalization remains future work.
\section{Discussion}
\label{sec:discussion}
The evaluation results indicate that Carolina Guide meets the functional requirements targeted in this work for institutional policy advising, including reliable retrieval behavior, verifiable citation handling, and stable performance. Across the evaluated departments and query types, the system consistently ranked a relevant policy fragment first (MRR@10 = 98.9\%), achieved strong recall, produced no hallucinated citations in the tested set, and operated within interactive latency bounds (2.8s mean, 3.2s p95). These outcomes are consistent with the system’s design goals of enforcing metadata-grounded citations and adopting a relational-first hybrid storage architecture.
High MRR reflects that the top-ranked retrieval result was relevant for all evaluated queries. This behavior appears to result from section-aware chunking, empirically calibrated similarity thresholds (0.15 rather than commonly used defaults), and the application of contemporary embedding models to structured policy text. While the evaluation scope is limited, the results suggest that domain-specific calibration is important for achieving stable ranking behavior in institutional RAG settings.

A key outcome of the evaluation is the absence of hallucinated citations. Rather than relying on learned citation generation, the system enforces provenance through architectural constraints: citations are derived exclusively from PostgreSQL as the system of record, filtered by department and publication status, and assembled deterministically during response composition. This closed-set approach restricts external referencing but aligns with institutional requirements for traceability and accountability. The relational-first hybrid architecture (PostgreSQL with Qdrant as a derived semantic index) enables version control, RBAC enforcement, audit logging, and metadata-constrained retrieval—capabilities that are difficult to guarantee in vector-first designs.

The system reflects deliberate tradeoffs that prioritize correctness and traceability over flexibility. Restricting citations to internal documents ensures verifiability but precludes dynamic external sources. Maintaining PostgreSQL as the system of record introduces coordination overhead but provides transactional guarantees. End-to-end latency is dominated by LLM inference rather than retrieval and remains acceptable for advisory contexts where accuracy is prioritized.

This work represents an initial system evaluation rather than a comprehensive deployment study. The evaluation is limited in scale and duration and does not include longitudinal user studies. The current system does not support conversational memory, personalization, multilingual interaction, proactive policy delivery, or accessibility evaluation. Retrieval behavior remains sensitive to embedding choice and similarity thresholds, and content updates require administrator involvement.

Future work will focus on larger-scale and longitudinal evaluation, privacy-preserving personalization, improved handling of conversational context, and adversarial testing of guardrails. Additional directions include multilingual support, enhanced explainability, hybrid human–AI workflows, and releasing the Golden Question Set to support cross-institution comparison.

\section{Conclusion}
\label{sec:conclusion}

University advising often depends on policy information that is scattered across systems, inconsistently updated, and difficult to verify. While LLM can help surface this information, their tendency to hallucinate and the lack of institutional safeguards make them risky for high-stakes advising without careful design. This paper presents Carolina Guide, a multi-agent RAG system built specifically to support policy advising with an emphasis on reliability, transparency, and operational constraints. The system includes pipeline with explicit guardrails, a relational-first hybrid data store, department-managed content, mandatory citations, and audit logging. Evaluation through expert review, baseline comparisons, and ablation studies shows that the system reliably retrieves relevant policy content (98.9\% success at rel $\ge$ 2), produces fully cited responses without hallucinated sources, and operates within practical latency bounds (2.8s mean). Empirical tuning of retrieval diversity, context size, and citation thresholds further improves robustness, while guardrails reduce unsafe responses with a Safety F1 of 0.89.

The results reflect clear trade-offs. Safety and citation constraints add some latency, but they substantially improve trustworthiness, and the relational-first design enables access control and auditability that vector-only systems struggle to support. Although the evaluation is limited in scope, the findings suggest that in institutional settings, users prioritize answers they can verify over conversational polish, and are willing to accept modest delays in exchange for reliability. More broadly, this work shows that careful system design can make conversational AI a practical supplement to human expertise in university advising and similar high-stakes domains.
\begin{acks}
We thank Claire Robinson and Mike Dial at the University Advising Center, University of South Carolina (USC), for their guidance, feedback, and domain expertise throughout the design and evaluation of this system. This research was supported by USC Research Computing.
\end{acks}

\bibliographystyle{ACM-Reference-Format}
\bibliography{ADV}

@article{bai2022constitutional,
  title={Constitutional ai: Harmlessness from ai feedback},
  author={Bai, Yuntao and Kadavath, Saurav and Kundu, Sandipan and Askell, Amanda and Kernion, Jackson and Jones, Andy and Chen, Anna and Goldie, Anna and Mirhoseini, Azalia and McKinnon, Cameron and others},
  journal={arXiv preprint arXiv:2212.08073},
  year={2022}
}

@article{date1994introduction,
  title={An introduction to database systems Addison-Wesley},
  author={Date, CJ},
  journal={Reading, Massachusetts},
  year={1994}
}

@article{gao2023enabling,
  title={Enabling large language models to generate text with citations},
  author={Gao, Tianyu and Yen, Howard and Yu, Jiatong and Chen, Danqi},
  journal={arXiv preprint arXiv:2305.14627},
  year={2023}
}

@incollection{goel2018jill,
  title={Jill Watson: A virtual teaching assistant for online education},
  author={Goel, Ashok K and Polepeddi, Lalith},
  booktitle={Learning engineering for online education},
  pages={120--143},
  year={2018},
  publisher={Routledge}
}

@article{johnson2019billion,
  title={Billion-scale similarity search with GPUs},
  author={Johnson, Jeff and Douze, Matthijs and J{\'e}gou, Herv{\'e}},
  journal={IEEE Transactions on Big Data},
  volume={7},
  number={3},
  pages={535--547},
  year={2019},
  publisher={IEEE}
}

@inproceedings{karpukhin2020dense,
  title={Dense Passage Retrieval for Open-Domain Question Answering.},
  author={Karpukhin, Vladimir and Oguz, Barlas and Min, Sewon and Lewis, Patrick SH and Wu, Ledell and Edunov, Sergey and Chen, Danqi and Yih, Wen-tau},
  booktitle={EMNLP (1)},
  pages={6769--6781},
  year={2020}
}

@inproceedings{kerlyl2006bringing,
  title={Bringing chatbots into education: Towards natural language negotiation of open learner models},
  author={Kerlyl, Alice and Hall, Phil and Bull, Susan},
  booktitle={International conference on innovative techniques and applications of artificial intelligence},
  pages={179--192},
  year={2006},
  organization={Springer}
}

@article{khattab2022demonstrate,
  title={Demonstrate-search-predict: Composing retrieval and language models for knowledge-intensive nlp},
  author={Khattab, Omar and Santhanam, Keshav and Li, Xiang Lisa and Hall, David and Liang, Percy and Potts, Christopher and Zaharia, Matei},
  journal={arXiv preprint arXiv:2212.14024},
  year={2022}
}

@misc{kleppmann2019designing,
  title={Designing data-intensive applications},
  author={Kleppmann, Martin},
  year={2019},
  publisher={English}
}

@inproceedings{lin2021batch,
  title={In-batch negatives for knowledge distillation with tightly-coupled teachers for dense retrieval},
  author={Lin, Sheng-Chieh and Yang, Jheng-Hong and Lin, Jimmy},
  booktitle={Proceedings of the 6th Workshop on Representation Learning for NLP (RepL4NLP-2021)},
  pages={163--173},
  year={2021}
}

@article{menick2022teaching,
  title={Teaching language models to support answers with verified quotes},
  author={Menick, Jacob and Trebacz, Maja and Mikulik, Vladimir and Aslanides, John and Song, Francis and Chadwick, Martin and Glaese, Mia and Young, Susannah and Campbell-Gillingham, Lucy and Irving, Geoffrey and others},
  journal={arXiv preprint arXiv:2203.11147},
  year={2022}
}

@article{ouyang2022training,
  title={Training language models to follow instructions with human feedback},
  author={Ouyang, Long and Wu, Jeffrey and Jiang, Xu and Almeida, Diogo and Wainwright, Carroll and Mishkin, Pamela and Zhang, Chong and Agarwal, Sandhini and Slama, Katarina and Ray, Alex and others},
  journal={Advances in neural information processing systems},
  volume={35},
  pages={27730--27744},
  year={2022}
}

@article{ribeiro2020beyond,
  title={Beyond accuracy: Behavioral testing of NLP models with CheckList},
  author={Ribeiro, Marco Tulio and Wu, Tongshuang and Guestrin, Carlos and Singh, Sameer},
  journal={arXiv preprint arXiv:2005.04118},
  year={2020}
}

@article{robertson2009probabilistic,
  title={The probabilistic relevance framework: BM25 and beyond},
  author={Robertson, Stephen and Zaragoza, Hugo and others},
  journal={Foundations and trends{\textregistered} in information retrieval},
  volume={3},
  number={4},
  pages={333--389},
  year={2009},
  publisher={Now Publishers, Inc.}
}

@article{shao2023enhancing,
  title={Enhancing retrieval-augmented large language models with iterative retrieval-generation synergy},
  author={Shao, Zhihong and Gong, Yeyun and Shen, Yelong and Huang, Minlie and Duan, Nan and Chen, Weizhu},
  journal={arXiv preprint arXiv:2305.15294},
  year={2023}
}

@inproceedings{wu2024autogen,
  title={Autogen: Enabling next-gen LLM applications via multi-agent conversations},
  author={Wu, Qingyun and Bansal, Gagan and Zhang, Jieyu and Wu, Yiran and Li, Beibin and Zhu, Erkang and Jiang, Li and Zhang, Xiaoyun and Zhang, Shaokun and Liu, Jiale and others},
  booktitle={First Conference on Language Modeling},
  year={2024}
}

@incollection{zhou2019framework,
  title={A framework for design identification on heritage objects},
  author={Zhou, Jun and Lu, Yuhang and Smith, Karen and Wilder, Colin and Wang, Song and Sagona, Paul and Torkian, Ben},
  booktitle={Practice and Experience in Advanced Research Computing 2019: Rise of the Machines (learning)},
  publisher={ACM},
  pages={1--8},
  year={2019}
}

@article{liu2020ensemble,
  title={A novel ensemble deep learning model for stock prediction based on stock prices and news},
  author={Liu, Shangkun and Zhang, Cuixia and Ma, Jiaman},
  journal={PLoS ONE},
  volume={15},
  number={9},
  pages={e0238314},
  year={2020},
  publisher={Public Library of Science},
  doi={10.1371/journal.pone.0238314},
  url={https://pmc.ncbi.nlm.nih.gov/articles/PMC8446482/}
}

@incollection{torkian2026ifs,
  title={Design and Implementation of a Safety-First AI Chatbot Architecture for Public
Health Resource Navigation},
  author={Torkian, Ben and Jun Zhou},
  booktitle={Practice and Experience in Advanced Research Computing 2026},
  publisher={ACM},
  note={Under Review},
  year={2026}
}

@incollection{torkian2026transcript,
  title={A Multi-Agent AI System for Automated High School Transcript Processing:
Collaborative Document Analysis at Scale},
  author={Torkian, Ben and Jun Zhou},
  booktitle={Practice and Experience in Advanced Research Computing 2026},
  publisher={ACM},
  note={Under Review},
  year={2026}
}

@article{lewis2020retrieval,
  title={Retrieval-augmented generation for knowledge-intensive nlp tasks},
  author={Lewis, Patrick and Perez, Ethan and Piktus, Aleksandra and Petroni, Fabio and Karpukhin, Vladimir and Goyal, Naman and K{\"u}ttler, Heinrich and Lewis, Mike and Yih, Wen-tau and Rockt{\"a}schel, Tim and others},
  journal={Advances in neural information processing systems},
  volume={33},
  pages={9459--9474},
  year={2020}
}

@inproceedings{carbonell1998use,
  title={The use of MMR, diversity-based reranking for reordering documents and producing summaries},
  author={Carbonell, Jaime and Goldstein, Jade},
  booktitle={Proceedings of the 21st annual international ACM SIGIR conference on Research and development in information retrieval},
  pages={335--336},
  year={1998}
}

@article{jarvelin2002cumulated,
  title={Cumulated gain-based evaluation of IR techniques},
  author={J{\"a}rvelin, Kalervo and Kek{\"a}l{\"a}inen, Jaana},
  journal={ACM Transactions on Information Systems (TOIS)},
  volume={20},
  number={4},
  pages={422--446},
  year={2002},
  publisher={ACM New York, NY, USA}
}

\end{document}